\documentclass[%
 prl,
 jmp,%
 amsmath,amssymb,
 floatfix,
reprint,%
 showpacs,%
]{revtex4-1}

\usepackage{comment}
\usepackage{amssymb}
\usepackage{amsmath}
\usepackage{mathtools}
\usepackage{relsize}
\usepackage{array}
\usepackage{bm}
\usepackage{graphicx}
\usepackage{bigints}
\usepackage{ifthen}
\usepackage{multirow}
\usepackage{threeparttable}
\usepackage{rotating}
\usepackage{hyperref}
\usepackage{xcolor}
\usepackage{calrsfs}
\usepackage{empheq}
\usepackage[normalem]{ulem}
\usepackage{media9}
\usepackage[margin=0pt,font=small,labelfont=normalfont,labelsep=period,format=plain,justification=centerlast]{caption}
\usepackage{xargs}   

\usepackage[draft,footnote,marginclue,innerlayout=inline]{fixme}
\fxloadlayouts{marginnote}
\fxsetup{theme=color,mode=multiuser}
\FXRegisterAuthor{V}{anV}{{\color{red}{VG}}}

\DeclareGraphicsExtensions{.pdf,.png,.eps}
\graphicspath{{figs/}{./}}
\usepackage{cleveref}
\crefname{equation}{}{}
\Crefname{equation}{}{}
\crefname{figure}{Fig.~\!\!\!}{Figs.~\!\!\!}
\Crefname{figure}{Fig.~\!\!\!}{Figs.~\!\!\!}

\newcommand{\paren}[1]{\left( #1 \right)}

\newcommand{\lims}[1]{\left[ #1 \right]}

\usepackage[colorinlistoftodos,prependcaption,textsize=tiny]{todonotes}
\newcommandx{\lrfnote}[2][1=]{\todo[linecolor=purple,backgroundcolor=purple!25,bordercolor=purple,#1]{#2}}
\newcommandx{\trnote}[2][1=]{\todo[linecolor=red,backgroundcolor=red!25,bordercolor=red,#1]{TR: #2}}
\newcommandx{\dbnote}[2][1=]{\todo[linecolor=blue,backgroundcolor=blue!25,bordercolor=blue,#1]{#2}}
\newcommandx{\lwnote}[2][1=]{\todo[linecolor=olive,backgroundcolor=olive!25,bordercolor=olive,#1]{#2}}
\newcommandx{\thiswillnotshow}[2][1=]{\todo[disable,#1]{#2}}

\usepackage{tikz}
\usetikzlibrary{intersections,arrows,decorations,decorations.pathmorphing,calc}
\usepackage{pgfplots}
\pgfplotsset{compat=newest}
\pgfplotsset{plot coordinates/math parser=false}

\newcommand{\der}[2]{\frac{d {#1}}{d {#2}}}
\newcommand{\derp}[2]{\frac{\partial {#1}}{\partial {#2}}}

\makeatletter
\let\jnl@style=\rmfamily
\def\ref@jnl#1{{\jnl@style#1}}
\@ifundefined{aap}{\newcommand\aap{\ref@jnl{A\&A}}}{}%
\@ifundefined{apj}{\newcommand\apj{\ref@jnl{ApJ}}}{}%
\@ifundefined{apjl}{\newcommand\apjl{\ref@jnl{Astrophys.~J.~Lett.}}}{}%
\@ifundefined{ssr}{\newcommand\ssr{\ref@jnl{Space~Sci.~Rev.}}}{}%
\@ifundefined{jgr}{\newcommand\jgr{\ref@jnl{J.~Geophys.~Res.}}}{}%
\@ifundefined{jatp}{\newcommand\jatp{\ref@jnl{J.~Atmos.~Terr.~Phys.}}}{}%
\@ifundefined{solphys}{\newcommand\solphys{\ref@jnl{Sol.~Phys.}}}{}%
\@ifundefined{pp}{\newcommand\pp{\ref@jnl{Phys.~Plasmas}}}{}%
\@ifundefined{mnras}{\newcommand\mnras{\ref@jnl{MNRAS}}}{}%
\@ifundefined{apjs}{\newcommand\apjs{\ref@jnl{ApJS}}}{}%
\@ifundefined{azh}{\newcommand\azh{\ref@jnl{AZh}}}{}%
\catcode`\@=11

\def\beq#1\eeq{\begin{equation}#1\end{equation}}
\def\bes#1\ees{\begin{subequations}#1\end{subequations}}
\def\bea#1\eea{\begin{align}#1\end{align}}
\def\n{\nonumber\\}

\newcommand{\mvec}[1]{\bm{#1}}

\usepackage{scalerel}

\allowdisplaybreaks

\begin{document}

\title{Collective Synchronous Spiking in a Brain Network of Coupled
    Nonlinear Oscillators}

\author{Vitaly L. Galinsky}
\email{vit@ucsd.edu}
\affiliation{Center for Scientific Computation in Imaging,
University of California at San Diego, La Jolla, California 92037-0854, USA}
\author{Lawrence R. Frank}
\email{lfrank@ucsd.edu}
\affiliation{Center for Scientific Computation in Imaging,
University of California at San Diego, La Jolla, California 92037-0854, USA}
\affiliation{
Center for Functional MRI,
University of California at San Diego, La Jolla, California 92037-0677, USA}

\date{\today}

\begin{abstract}

A network of propagating nonlinear oscillatory modes (waves) in the
human brain is shown to generate collectively synchronized spiking
activity (hypersynchronous spiking) when both amplitude and phase
coupling between modes are taken into account. The nonlinear behaviour
of the modes participating in the network are the result of the
nonresonant dynamics of weakly evanescent cortical waves that, as
shown recently, adhere to an inverse frequency-wave number dispersion
relation when propagating through an inhomogeneous anisotropic media
characteristic of the brain cortex. This description provides a
missing link between simplistic models of synchronization in networks
of small amplitude phase coupled oscillators and in networks built
with various empirically fitted models of pulse or amplitude coupled
spiking neurons. Overall the phase-amplitude coupling mechanism
presented in the Letter shows significantly more efficient
synchronization compared to current standard approaches and
demonstrates an emergence of collective synchronized spiking from
subthreshold oscillations that neither phase nor amplitude coupling
alone are capable of explaining.

\end{abstract}


\maketitle

Brain electromagnetic activity shows an abundance of oscillatory
patterns across a wide range of spatial and temporal scales making the
question of their interaction and synchronization an important issue
that has been widely discussed in the literature
\citep{buzsaki2006rhythms,*Gerstner:2014:NDS:2635959}.  All the
typical approaches to the question of synchronization of multiple
interconnected (neural) networks can be divided into two big
groups. The first approach works with a network comprised of multiple
harmonic oscillators in a small (and constant) amplitude limit
\citep{1975LNP....39..420K,Kuramoto2002,Kuramoto2003,2004PhRvL..93q4102A,
  2005RvMP...77..137A,pmid32872818,pmid30341395}. The second approach
studies networks of multiple empirical nonlinear elements,
e.g., various variants and simplifications of heavily over-fitted
Hodgkin-Huxley neurons \citep{pmid12991237}, including a multitude of
{\it ad hoc} ``integrate and fire'' (IF) neuron models
\citep{pmid19431309,Nagumo1962,pmid7260316,pmid18244602,Gerstner:2014:NDS:2635959,pmid33192427,pmid33288909}. Both
approaches have their own advantages and drawbacks, but the main
problem is that although they are often considered as complementary,
they are not just incompatible, they are based on contradictory
assumptions. The first approach assumes that a phase of oscillations
is the cornerstone of the synchronization process and all the
importance of the amplitudes is second to none, hence the oscillating
amplitudes can be safely assumed to be constant. The second approach
on the contrary deems phase information to be nothing more then a
subthreshold noise that can be safely thrown away completely and just
accumulates amplitudes of the arriving spikes or pulses when
processing an input from multiple IF network members hoping that the
discarded individual subspike phase information will magically be
resurrected in a new form as a population averaged synchronous phase.

Recently it was experimentally discovered that long-range (at the
distance of 60 mm apart or even more) correlations exist in human
cortex in the 100--400 Hz frequency range \citep{pmid33097714}. This
frequency range corresponds to 2.5--10 ms signal periods, i.e.,~it is at
or even below the duration of a single neuronal spike and, hence, it
requires coherent spiking at the single neuron level and not just some
average population synchrony. Neither of the two methods of phase
coupled harmonic oscillators or pulse coupled IF neurons are capable
of explaining this level of spiking synchrony as it is acknowledged in
the literature that ``there is no known mechanism through which the
spikes of multiple neurons could be synchronized so precisely''
\citep{pmid19793873}.

The recently developed theory of weakly evanescent cortical waves
(WETCOWs) \citep{Galinsky:2020a,*Galinsky:2020b} provides a mechanism
appropriate for the explanation of long-range high frequency correlations
and multiple wave modes and spikes synchronization up to and below
a single spike duration (hypersynchronous spiking) that follows directly
from linear and nonlinear properties of wave modes. The linear wave
dispersion relation predicts an inverse frequency--wave number
dependence, hence the correlations at the highest frequencies should
manifest themselves at the longest spatial scales in agreement with
properties reported in Ref.~\citep{pmid33097714}. The model of nonlinear
interactions of those wave modes provides a mechanism of generation of
spiking activity from their collective input, hence it is appropriate
for explaining the physical origin of hypersynchronous spiking.

In this Letter we show that the nonlinear model of brain wave modes
developed in Ref.~\citep{Galinsky:2020a,*Galinsky:2020b} can be
reformulated using a simple but general Hamiltonian form that includes
all possible nonlinear interaction at the lowest order of
nonlinearity. Dynamical equations defined by this wave Hamiltonian
reproduce oscillatory activity from the linear (harmonic) wave regime to
nonlinear spiking modes. Extending the Hamiltonian to include a
pairwise coupling appropriate for a network of multiple nonlinear wave
modes results in amplitude and phase coupled nonlinear equations that
show more efficient synchronization comparing to just phase coupling
alone. For sufficiently strong coupling the spiking activity that
emerges at a different part of the network from the small amplitude (below
spiking detection or subthreshold level) oscillations is synchronized
not just in some averaged (spiking population) sense but at a single
spike level. This amplitude and phase coupling approach thus provides a
missing bridge between phase only coupling models of harmonic
oscillator networks and amplitude (pulse) coupling models of IF
neurons.

A nonlinear Hamiltonian form for an anharmonic oscillatory mode with a
complex amplitude $a$ in the lowest order of nonlinearity can be
assumed to have the expression 
\vspace*{-2pt}
\begin{align}
\label{eq:H}
H^s(a,a^\dag) &= 
\Gamma a a^\dag\! + a a^\dag\! \left[\beta_a a + \beta_{a^\dag} a^\dag\! -
  2\alpha \left(a a^\dag \right)^{1/2}\right] 
\end{align}
\vspace*{-2pt}
where $a$ is a complex oscillation amplitude and $a^\dag$ is its
conjugate. The first term $\Gamma a a^\dag$ denotes the harmonic
(quadratic) part of the Hamiltonian with the complex valued frequency
$\Gamma=i\omega +\gamma$ that includes both a pure oscillatory
frequency $\omega$ and a possible weakly excitation or damping rate
$\gamma$. Because of the presence of this $\gamma\ne0$, the conjugate
$^\dag$ does not denote just the complex conjugate but more
generally it also describes the growth or decay of the amplitude as a
result of the presence of excitation or damping, e.g., for the growing
oscillatory amplitude $a \sim e^{\gamma t + i\omega t}$ the conjugate
will describe the correspondent decaying part as $a^\dag \sim
e^{-\gamma t -i\omega t}$.

The second anharmonic term (that is supposed to be cubic in the lowest
order of nonlinearity) can be considered to include a product of the
harmonic term $a a^\dag$ and linear (in $|a|$) term that can be
expressed in the most general form as $\beta_a a + \beta_{a^\dag}
a^\dag - 2\alpha \left(a a^\dag\right)^{1/2}$ (where $\alpha$,
$\beta_a$ and $\beta_{a^\dag}$ are the complex valued strengths of
nonlinearity, and in general we do not assume that the Hamiltonian is
in self-adjoint form, hence in general $\beta_{a}^\dag \ne
\beta_{a^\dag}$). The terms with either $a^3$ or ${a^\dag}^3$ do not
appear in the Hamiltonian because for propagating waves they do not
satisfy the wave number resonance conditions
\citep{book:971420,*book:787941,Galinsky:2020a,*Galinsky:2020b}.

It is worth noting that Eq.~\cref{eq:H} is very general and can
be used for a description of various anharmonic oscillatory physical
processes valid to the lowest order of amplitude nonlinearity.  In
this Letter we have used it as a reformulation of weakly evanescent
brain wave modes whose linear and nonlinear physical properties were
presented in Ref.~\citep{Galinsky:2020a,*Galinsky:2020b}.  In this case,
the nonlinear terms can be identified with the nonresonant
interactions of linear wave modes propagating in
either the same ($\beta_a$) or the opposite ($\beta_{a^\dag}$)
directions, or with the interactions with phase averaged
nonpropagating modes ($\alpha$).

An equation for the nonlinear oscillatory amplitude $a$ then can be
expressed as a derivative of the Hamiltonian form 
\begin{align}
\label{eq:a}
\der{a}{t}=\derp{H^s}{a^\dag}\equiv \Gamma a + 2 \beta_{a^\dag} a
a^\dag + \beta_a a^2 - 3\alpha a (a a^\dag )^{1/2}.
\end{align}
Substituting $a=\tilde{a}e^{i\omega t}$,  $a^\dag=\tilde{a}^\dag
e^{-i\omega t}$, $\beta_a=\tilde{\beta_a}e^{-i\delta_a}$,
$\beta_{a^\dag}=1/2 \tilde{\beta}_{a^\dag} e^{i\delta_{a^\dag}}$,
and $\alpha=1/3\tilde{\alpha}$, dropping
the tilde, this can be rewritten as 
\begin{align}
\label{eq:a1}
\der{a}{t}&=\gamma a + \beta_{a^\dag} a a^\dag e^{-i(\omega
t-\delta_{a^\dag})} + \beta_a a^2 e^{i(\omega t-\delta_a) }
- \alpha a (a a^\dag )^{1/2}.
\end{align}

Equation \Cref{eq:a1} is similar (up to the choice of the constants) to
Eq.~(29) of Ref.~\citep{Galinsky:2020a,*Galinsky:2020b} but has an
additional oscillatory term that was not included in
Ref.~\citep{Galinsky:2020a,*Galinsky:2020b}.  Substituting in Eq.~\cref{eq:a}
$a=\tilde{a}e^{\gamma t + i\omega t}$ and $a^\dag=\tilde{a}^\dag e^{-\gamma
t - i\omega t}$ instead of just the oscillatory complex exponents
makes it is clear that those two terms represent the damped
$\beta_{a^\dag} \tilde{a} \tilde{a}^\dag e^{-\gamma t -i(\omega
t-\delta_{a^\dag})}$ and the growing $\beta_a \tilde{a}^2 e^{\gamma
t + i(\omega t-\delta_a)}$ parts of the nonlinear input and the
spiking solutions can be obtained even when the damped term is neglected.
Nevertheless, to analyze the more general case we will keep both
of those terms and will show later that it is the asymmetry between
those terms that provides an explanation for the presence of the phase
difference that plays an important role in collective synchronization and
hypersynchronous spiking.

\begin{figure}[!tbh] \centering
\includegraphics[width=1.0\columnwidth]{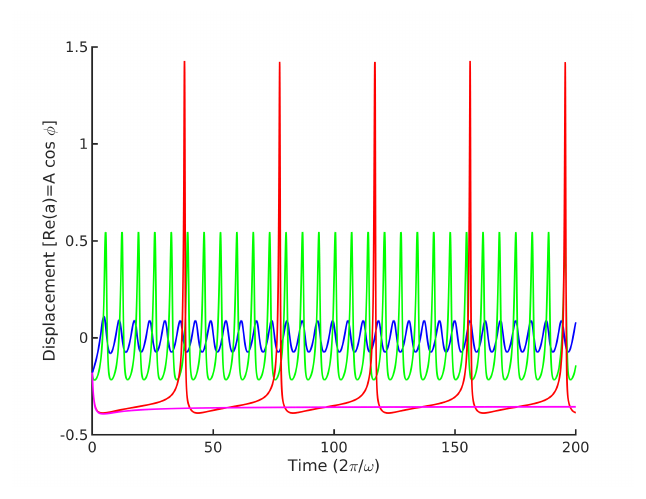}
\caption[]{Linear and nonlinear oscillations, spiking, and
  nonoscillatory regime of the solution of Eqs.~\cref{eq:A0,eq:B0} at
  different levels of excitation
  ($\gamma=0.25,1,1.98$ and 2:blue,green,red, and magenta). Time is
  in the units of $2\pi/\omega$ and the amplitude $a$ is in the
  arbitrary units. The rest of the parameters are the same for all
  plots: $\omega=1$, $\beta_a=2$, $\beta_{a^\dag}=1$, $\alpha=3$,
  $\delta_{a^\dag}=\pi/4$, and $\delta_a=-\pi/4$.}
\label{fig:spiking}
\end{figure}

Splitting Eq.~\cref{eq:a1} into an amplitude-phase pair of
equations using $a=Ae^{i\phi}$ and assuming $\beta_a$,
$\beta_{a^\dag}$ and $\alpha$ to be real gives equations
\begin{align}
\label{eq:A0}
\der{A}{t} &=
\gamma A + A^2 \paren{\beta_{a^\dag}  \cos\Omega_{a^\dag}
+\beta_{a} \cos\Omega_a - \alpha }, 
\\
\label{eq:B0}
\der{\phi}{t} &= A \paren{-\beta_{a^\dag} \sin\Omega_{a^\dag}
+ \beta_{a} \sin\Omega_a}, \
\end{align}
where $\Omega_a \equiv \Omega-\delta_{a}$, $\Omega_{a^\dag} \equiv
\Omega-\delta_{a^\dag}$, and $\Omega \equiv \phi+\omega t$.  These are
similar to Eqs.~(31) and (32) of
Ref.~\citep{Galinsky:2020a,*Galinsky:2020b} and show the same solution
behaviour, i.e., transition from linear to nonlinear oscillation to
spiking to nonoscillatory regime with an increase of excitation
$\gamma$ as it is evident from \cref{fig:spiking} \citep{Sync-Supp}.

It should be noted that the asymmetries (i.e.,~the differences
between the rising and the falling edges of the spikes)
evident in the
spiking solutions obtained both in \cref{fig:spiking}
and in Ref.~\citep{Galinsky:2020a,*Galinsky:2020b} (where additional
asymmetric waveforms were presented)
correspond to a special
case that does not occur if a self-adjoint symmetry of the
Hamiltonian form Eq.~\cref{eq:H} is assumed.  In a self-adjoint form
(i.e., for $\beta_{a^\dag}=\beta_a^\dag$) the Hamiltonian in
Eq.~\cref{eq:H} can be alternatively expressed as
\begin{align}
\label{eq:H1}
H^s(a,a^\dag ) = \Gamma a a^\dag + a a^\dag \lims{F_{+}^2 + F_{-}^2}
\end{align}
where $F_+$ and $F_-$ are defined as
\begin{align}
\label{eq:F}
F_{+} &= \sqrt{\beta_{+}^{\dag} a^\dag } + \sqrt{\beta_{+}^{\phantom{\dag}}
  a\phantom{^\dag}} \\
F_{-} &= \sqrt{\beta_{-}^{\dag}
  a^\dag } - \sqrt{\beta_{-}^{\phantom{\dag}} a\phantom{^\dag}}
\end{align}
and correspond to a symmetric ($F_+^\dag=F_+$) and antisymmetric ($F_-^\dag=-F_-$) 
combinations of the complex oscillatory modes $a$ and $a^\dag$, with complex parameters
$\beta_-$ and $\beta_+$ satisfying the relations
\begin{align}
\beta_a =\beta_-+\beta_+ ,\quad
\alpha = |\beta_-|-|\beta_+|
\end{align}

The parameters when the spiking solutions were obtained both in
\cref{fig:spiking} and in Red.~\citep{Galinsky:2020a,*Galinsky:2020b}
correspond to $\alpha>0$, $\Rightarrow$ $|\beta_-|>|\beta_+|$,
$\Rightarrow$ $|F_{-}|>|F_{+}|$ and the opposite case
$|F_{-}|<|F_{+}|$ results in diverging solutions.  The fixed
difference in the power between $-$ and $+$ modes can be accompanied
by various phase shifts, therefore we introduced those different
$\delta_{a}$ and $\delta_{a^\dag}$ phases in Eq.~\cref{eq:a1}. As it was
shown in Ref.~\citep{Galinsky:2020a,*Galinsky:2020b} it is this difference
in phases that is responsible for the asymmetries in the shape of the
spikes. Here we have augmented the theory presented in
Ref.~\citep{Galinsky:2020a,*Galinsky:2020b} to account for these phase
differences as a consequence of symmetry considerations.  Though the
asymmetries in the spiking waveforms caused by this phase difference
might appear subtle, this is an important aspect of the theory as these
asymmetries are observed experimentally \citep{pmid19793873}.  but
have not been explained by any existing theory.  The effect on
synchronization, however, is not subtle at all, as we demonstrate
below.

These symmetry considerations create some (although very distant)
analogy with symmetry differences between bosons and fermions. The
antisymmetric form allows organization of the oscillations into highly
ordered spike sequences.  That is, it enforces in some sense the
single oscillation quanta-single state ``fermionic''-like selection
rule by only allowing generation of $a$ and $a^\dag$ pairs.  On the
contrary, the symmetric form results in accumulation of oscillations
in disordered divergent ``bosonic'' states.  However the important
point is that the observational fact of asymmetric spiking waveforms
suggests the existence of distinct symmetric and antisymmetric brain
wave states.  The origin of these is unknown.

As it was shown in Ref.~\citep{Galinsky:2020a,*Galinsky:2020b} the spiking
solution of Eqs.~(31) and (32) of
Ref.~\citep{Galinsky:2020a,*Galinsky:2020b} (that are similar to the system
Eqs.~\cref{eq:A0,eq:B0} of this Letter) appears near the critical point
where the oscillatory state undergoes bifurcation and transitions to a
nonoscillatory regime as $\gamma$ reaches the value above the critical
point. Assuming as in Ref.~\citep{Galinsky:2020a,*Galinsky:2020b} that the
nonoscillatory regime requires that $dA/dt \rightarrow 0$ and
$d\phi/dt \rightarrow -\omega$ (or $\phi \rightarrow -\omega t +
\phi_0$, with $\phi_0$ being some arbitrary phase) as
$t\rightarrow\infty$, it is easy to see that the
relative contribution of the excitation or damping in the amplitude
exponents is proportional to $\gamma/\omega$ which is given by
\begin{align}
\gamma / \omega = \frac{\alpha - \beta_{a^\dag}\cos(\phi_0-\delta_{a^\dag})
 -\beta_a\cos (\phi_0-\delta_a)}
{\beta_{a^\dag}\sin(\phi_0-\delta_{a^\dag}) -\beta_a\sin (\phi_0-\delta_a)},
\end{align}
and for the parameters of \cref{fig:spiking} this defines a range
$-2/5<\gamma<2$ with bifurcation at $\gamma_{crit} = 2$ in
excellent agreement with the transition from the oscillatory to
nonoscillatory regime obtained in the numerical solutions of
\cref{fig:spiking}.

The Hamiltonian form for a network of multiple coupled ``fermionic''
state oscillators can then be written as 
\begin{align}
\label{eq:HM}
\hskip -6pt
H(\mvec{a},\mvec{a}^\dag )\! &=\! \sum_i\! 
\left[\!
H^s(a_i,a_i^\dag)\! +\!
\sum_{j\ne i}\!
\left(
a_i r_{ij} a_j^\dag + a_i^\dag r^*_{ij} a_j
\right)\!\right]
\end{align}
where $\mvec{a} \equiv \{a_i\}$ and $r_{ij}=R_{ij}e^{i\Delta_{ij}}$ is
the complex network adjacency matrix with $R_{ij}$ providing the
coupling power and $\Delta_{ij}$ taking into account any possible
differences in phase between network nodes.  And an equation for the
complex amplitude $a_i$ (again after a substitution of 
$a_i=\tilde{a_i}e^{i\omega_i t}$,  $a_i^\dag=\tilde{a_i}^\dag
e^{-i\omega_i t}$, $\beta_a=\tilde{\beta_a}e^{-i\delta_a}$,
$\beta_{a^\dag}=1/2 \tilde{\beta}_{a^\dag} e^{i\delta_{a^\dag}}$,
$\alpha=1/3\tilde{\alpha}$ and dropping the tilde)
\begin{align}
\label{eq:a2}
\der{a_i}{t} &= 
\gamma_i a_i + \beta_{a^\dag} a_i a_i^\dag e^{-i(\omega_i t-\delta_{a^\dag})} + 
\beta_{a} a_i^2 e^{i(\omega_i t-\delta_a)}
\n&
- \alpha a_i (a_i a_i^\dag)^{1/2}
+\sum_{j\ne i}
r^*_{ij} a_j e^{i(\omega_j-\omega_i)t}
\end{align}
now includes the coupling term.
That gives for the amplitude $A_i$ and the phase $\phi_i$ a set of coupled equations
\begin{align}
\label{eq:AM0}
\der{A_i}{t} &=
\gamma_i A_i + A_i^2 \left(\beta_{a^\dag}\cos \Omega_{a^\dag}^{i} +
\beta_{a} \cos \Omega_a^{i} - \alpha\right) \n
&+
\sum_{j\ne i} R_{ij}A_j\cos(\Omega^{j} - \Omega^{i} - \Delta_{ij}),
\\
\label{eq:BM0}
A_i\der{\phi_i}{t} &= - A_i^2 \left(\beta_{a^\dag}\sin \Omega_{a^\dag}^{i} -
\beta_{a} \sin \Omega_a^{i}  \right) \n
&+
\sum_{j\ne i} R_{ij}A_j\sin(\Omega^{j} - \Omega^{i} - \Delta_{ij}),
\end{align}
where $\Omega_a^{i} \equiv \Omega^{i}-\delta_a$, $\Omega_{a^\dag}^{i}
\equiv \Omega^{i}-\delta_{a^\dag}$, $\Omega^{i} \equiv
\phi_{i}+\omega_i t$, and the coupling terms are dependent upon
$\Omega^{j} - \Omega^{i} = (\phi_{j} - \phi_{i}) + (\omega_{j}
-\omega_{i})t$.  In the small (and constant) amplitude limit ($A_i=$
const) this set of equations turns into a set of phase coupled
harmonic oscillators with a familiar $\sin(\phi_j-\phi_i \cdots)$ form
of phase coupling.  But in its general form Eqs.~\cref{eq:AM0,eq:BM0}
include also the phase dependent coupling of amplitudes
[$\cos(\phi_j-\phi_i \cdots)$] that dynamically defines if the input
from $j$ to $i$ will either play excitatory ($|\phi_j-\phi_i
+\cdots|<\pi/2$) or inhibitory ($|\phi_j-\phi_i +\cdots|>\pi/2$) roles
(this is in addition to any phase shift introduced by the static
network attributed phase delay factors $\Delta_{ij}$).

\begin{figure}[!tbh] \centering
\includegraphics[width=0.9\columnwidth]{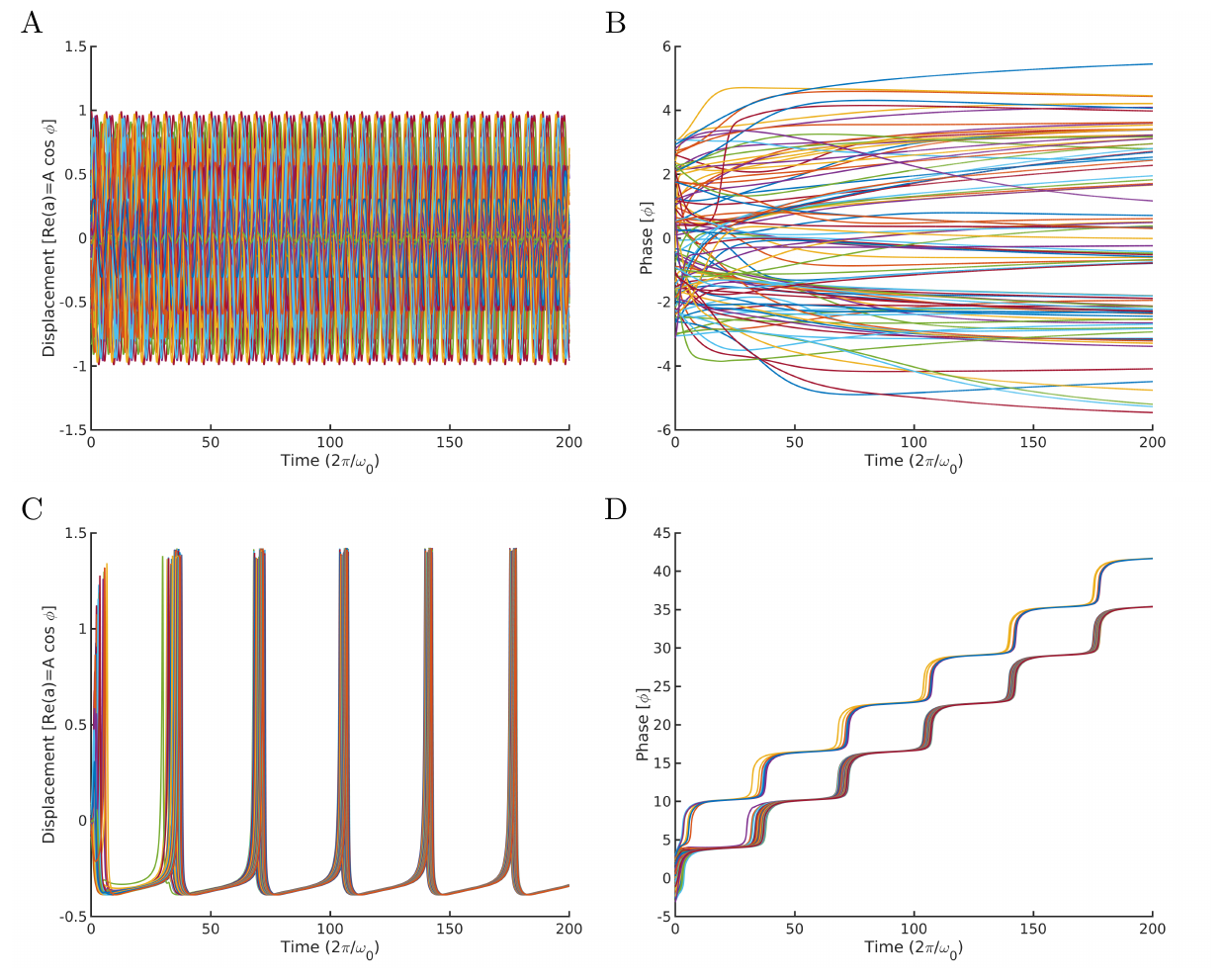}
\caption[]{Comparison of synchronization of phase-only constant
  amplitude harmonic oscillators [(A),(B)] with amplitude-phase coupling
  of the nonlinear Eqs.~\cref{eq:AM0,eq:BM0} model [(C),(D)].  Network of 100 ring
  connected oscillators ($R_{ij}=0.025$, $|i-j|\le 2$ and $R_{ij}=0$,
  $|i-j|>2$) shows only a transient weekly synchronized state
  (the order parameter $r$ = 0.32) in the
  phase-only system [(A),(B)] vs strongly synchronized spiking 
  (the order parameter $r$ = 0.9999 and still
  gradually increasing) in (C) and (D). Single frequency
  ($\omega\equiv\omega_0=1$) was used for both cases and
  $\beta_a=2$, $\beta_{a^\dag}=1$, $\alpha=3$,
  $\gamma=1.875$, $\delta_{a^\dag}=\pi/4$, $\delta_a=-\pi/4$
  were the parameters for the nonlinear Eqs.~\cref{eq:AM0,eq:BM0} system. 
}
\label{fig:f2}
\end{figure}

The relatively simple set of Eqs.~\cref{eq:AM0,eq:BM0} derived
from the simple but nevertheless general Hamiltonian form Eq.~\cref{eq:HM}
is capable of describing rich oscillatory and nonlinear dynamics as well
as more efficient synchronization compared to the phase-only coupled
system of harmonic oscillators even for the relatively weak
coupling. \Cref{fig:f2} shows a comparison of synchronization in a
weakly coupled network of phase-only constant amplitude harmonic
oscillators vs amplitude and phase coupling of nonlinear system
Eqs.~\cref{eq:AM0,eq:BM0} \citep{Sync-Supp}. The identical single frequency
$\omega=\omega_0=1$ harmonic oscillators coupled in a ring with just 4
nearest neighbors ($R_{ij}=0$, $|i-j|>2$) is still showing a transient
behaviour at $t=200$ with the order parameter $r=0.32$ 
[panels (A) and (B)] whereas the strongly
synchronized spiking with mean frequency $\overline{\omega}\sim
0.175\omega_0$ is formed as early as about $t=10$ for the nonlinear
amplitude and phase coupling system Eqs.~\cref{eq:AM0,eq:BM0} with the order 
parameter $r=0.9999$ [panels (C) and (D)].

\begin{figure}[!tbh] \centering
\includegraphics[width=0.9\columnwidth]{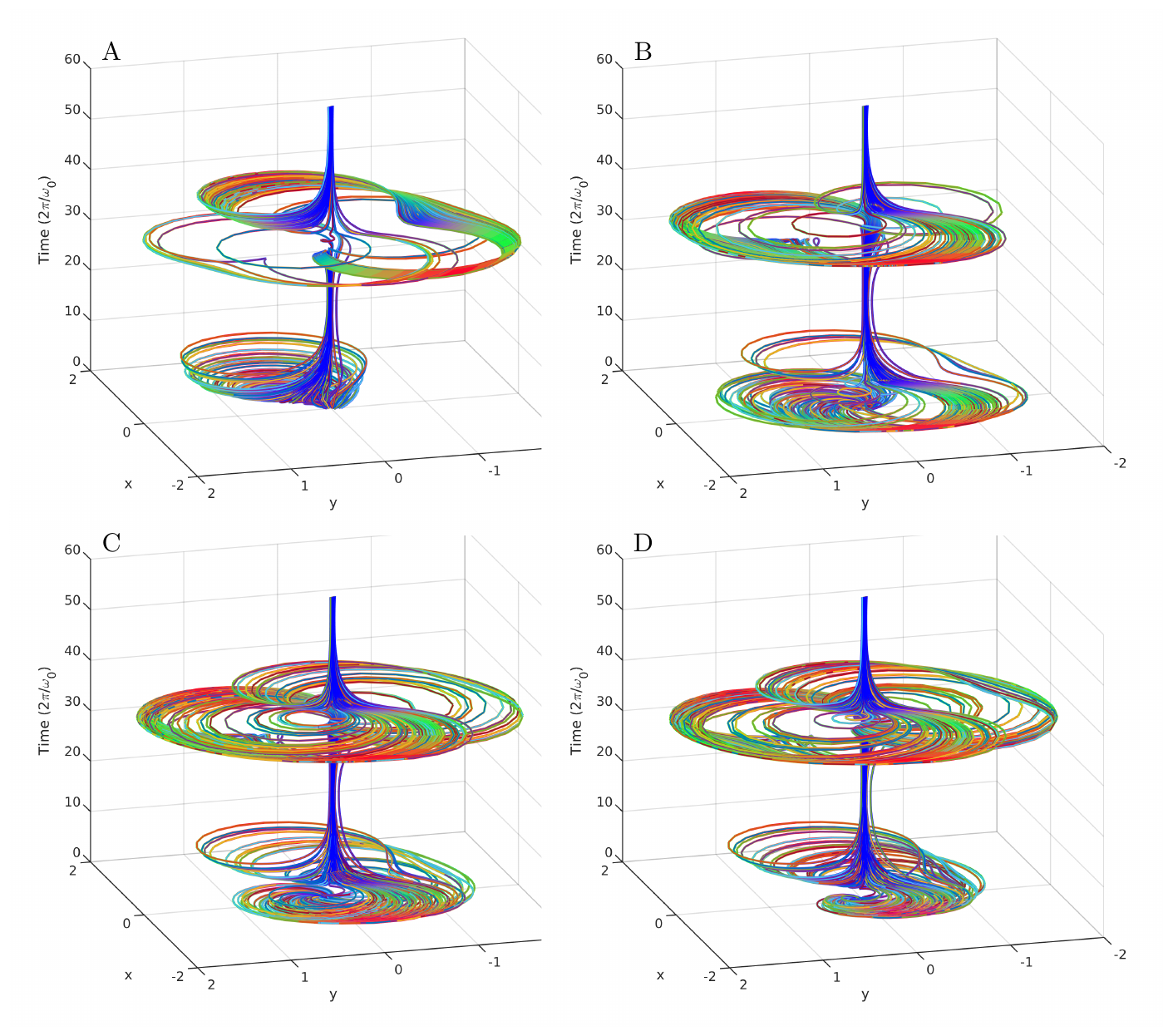}
\caption[]{Plot of the difference of complex amplitudes for a set of random
  network nodes $\{i\}$ with the amplitude of another random network node
  $k$ as a function of time ($x=Re[a_i(t)-a_k(t)]$,
  $y=Im[a_i(t)-a_k(t)]$).  The strong synchrony, represented
  by the vertical line at $(x,y)=(0,0)$ that emerges as $t$
  increases, is achieved during the first period $t<T=2\pi/\omega$
  of the oscillations.}
\label{fig:f3}
\end{figure}

The efficiency of the phase--amplitude form of the synchronization is
further illustrated in \cref{fig:f3} where differences of complex
amplitudes for a set of random network nodes with the amplitude of another
random node $k$ are plotted as functions of time (with $x$ and $y$ axes
corresponding to the real and the imaginary parts of $a_i-a_k$ and the
time advances are shown in $z$ direction). The strong spiking synchrony
is achieved as early as $t<T$ where $T=2\pi/\omega$ is the period of
the harmonic part of the oscillators.

\begin{figure}[!tbh] \centering
\includegraphics[width=0.9\columnwidth]{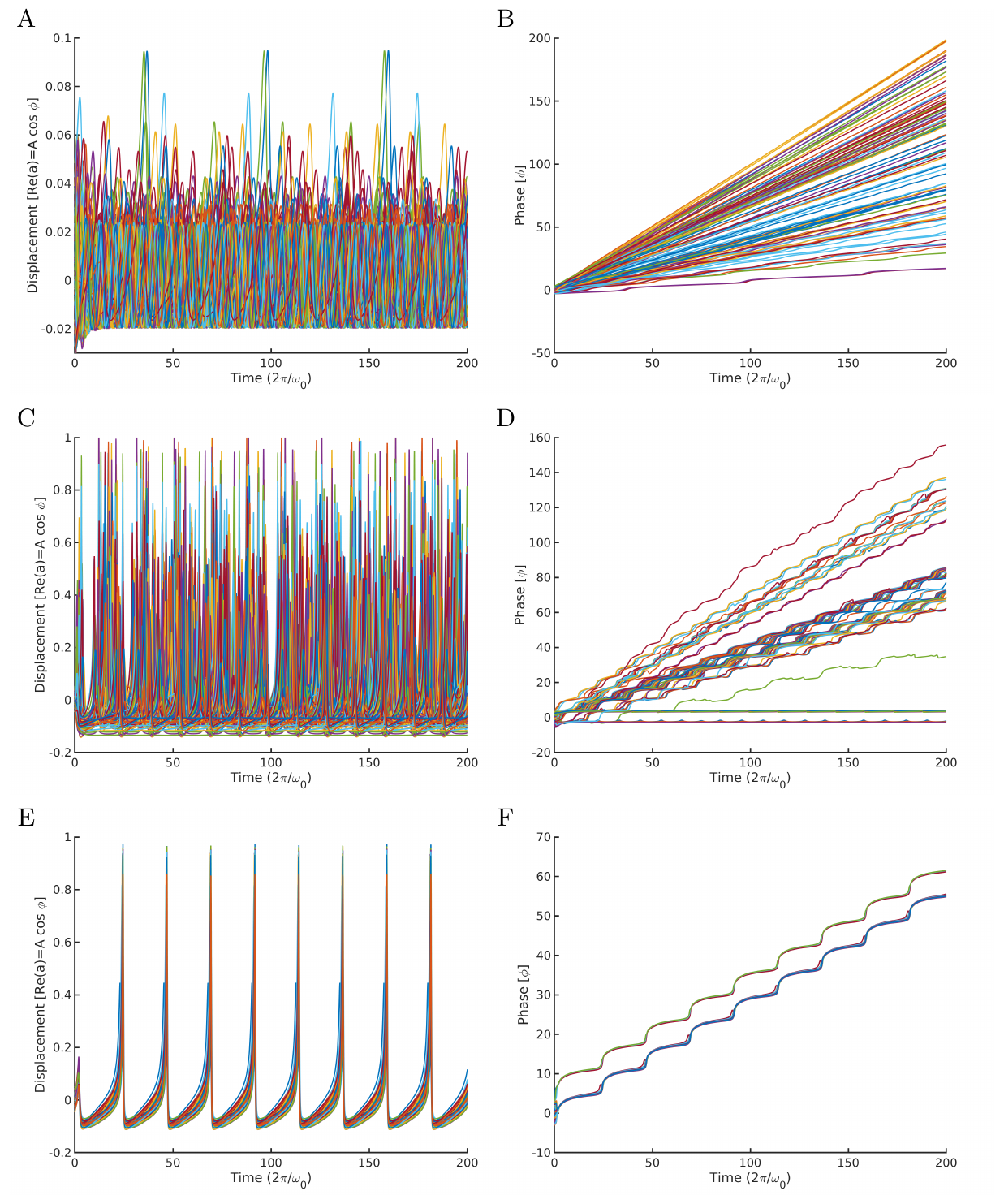}
\caption[]{Emergence of hypersynchronized spiking in a network of
  multiple frequency ($\omega_i<\omega_0\equiv 1$, $\omega_i$ is
  uniformly distributed between 0.1 and 1, $\overline{\omega_i}=0.54$,
  $\sigma_\omega=0.25$) weakly forced ($\gamma_i=0.25$) nonlinear
  oscillators.  (A) and (B) show low amplitude nonlinear
  oscillations of an uncoupled set. (C) and (D) show emergence of a relatively
  weakly synchronized state with three distinct frequency groups at
  roughly 0, $0.3$, and $0.6$ plus some single elements at intermediate
  frequencies for a nearest neighbor random coupling ($R_{ij}=0,
  |i-j|>1$, $\overline{R_{ij}}=1$, $\sigma_{R}=0.415$). (E) and (F) show
  the emergence of a strongly synchronized state with mean frequency
  at roughly $\sim 0.5$ and with the order parameter $r$=0.9882
  for all-to-all random coupling 
  $\overline{R_{ij}}=0.01$, $\sigma_{R}=0.0083$.}
\label{fig:f4}
\end{figure}

For a network with multiple individual frequencies
$\omega_i<\omega_0\equiv 1$ uniformly distributed between
0.1$\omega_0$ and $\omega_0$ and with only a weak forcing at
$\gamma_i=0.25$, that is not sufficiently strong for generating spikes
at individual uncoupled modes (as it can be seed from panels (A) and (B)
of \cref{fig:f4}, where all the amplitudes of oscillations are below
0.1 and the phases are clearly showing the expected spread from 0.1 to
1, the order parameter $r$=0.0792), 
the coupling triggers collectively synchronized spiking at
multiple frequencies [panels (C) and (D)] for a local nearest neighbor
coupling or at a single mean frequency [panels (E) and (F)] for a global
coupling
(i.e.~all-to-all or coupling of every node to all nodes in the network).

\begin{figure}[!tbh] \centering
\includegraphics[width=1.0\columnwidth]{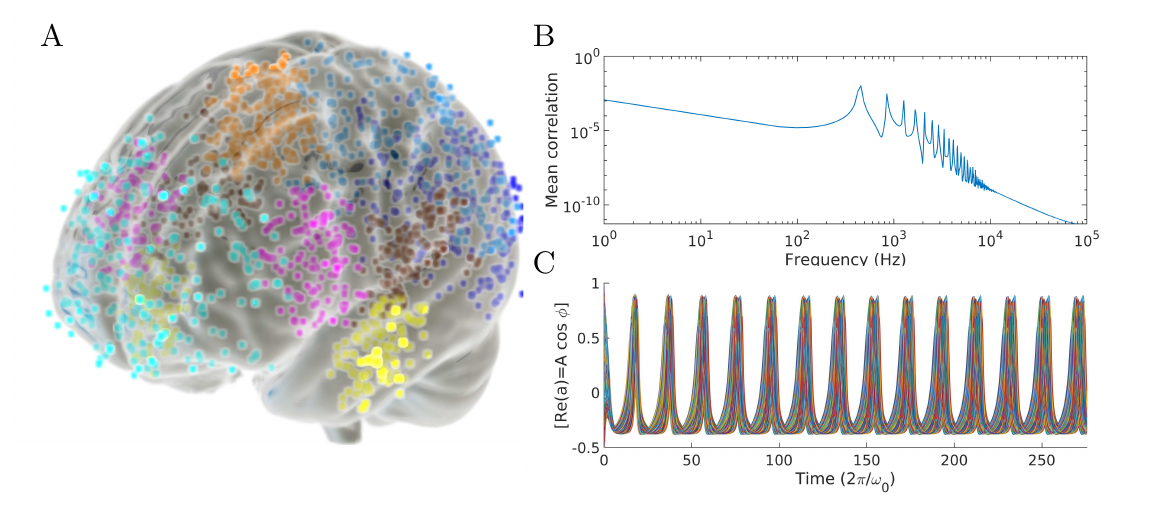}
\caption[]{(A) Network of $N=10032$ nodes split in 48 local cortical regions with 
every node connected to 10--11 nodes from every other region. (B) The plot of pairwise
correlations averaged across all nodes showing a peak at 300--500 Hz range
in agreement with Ref.~\citep{pmid33097714}. (C) Plot of synchronized oscillations.}
\label{fig:f5}
\end{figure}

This level of synchronization effectiveness is maintained as a 
number of network nodes $N$ increases and goes to infinity as can be
seen in \Cref{fig:f5} [panel (C)]. The network of \Cref{fig:f5}
includes the large number of nodes $N=10032$ split into 48 local
groups located in different
cortical regions [some of the nodes from several regions are shown in panel (A)]
with 
55 mm the mean distance between regions (and 61 mm the standard deviations).
Each node is connected to 500 different nodes with 10--11 nodes from every
region. The linear network frequencies are randomly distributed from 
1 Hz to 1 KHz. A plot of pairwise node correlations averaged across all pairs [shown in 
panel (B)] clearly demonstrates the peak between 300 and 500 Hz in agreement
with Ref.~\citep{pmid33097714}.


In conclusion, in this Letter we have presented a reformulation of the
nonlinear model of weakly evanescent cortical wave (WETCOW) modes
developed in Ref.~\citep{Galinsky:2020a,*Galinsky:2020b} into a Hamiltonian
form using a simple but general Hamiltonian representation that
includes all possible nonlinear interactions at the lowest order of
nonlinearity. Dynamical equations defined by this wave Hamiltonian
reproduce oscillatory activity from the linear (harmonic) wave regime to
nonlinear spiking modes. Extending the Hamiltonian to include a
pairwise coupling appropriate for a network of multiple nonlinear wave
modes results in amplitude and phase coupled nonlinear equations that
show more efficient synchronization comparing to just phase coupling
alone. For sufficiently strong coupling the spiking activity that
emerges at different parts of network from the small amplitude (below
spiking detection or subthreshold level) oscillations is synchronized
not just in some averaged (spiking population) sense but at a single
spike resolution or below. This amplitude and phase coupling approach thus
provides a missing link between phase only coupling models of harmonic
oscillator networks and amplitude (pulse) coupling models of IF
neurons, and has implications for understanding
experimentally observed synchronous behaviour in the human brain.

\begin{acknowledgments}
L.R.F.~and V.L.G.~were supported by NSF Grant No.~ACI-1550405, UCOP MRPI Grant No.~MRP17454755
and NIH Grant No.~R01 AG054049.
\end{acknowledgments}



\begin{thebibliography}{23}%
\makeatletter
\providecommand \@ifxundefined [1]{%
 \@ifx{#1\undefined}
}%
\providecommand \@ifnum [1]{%
 \ifnum #1\expandafter \@firstoftwo
 \else \expandafter \@secondoftwo
 \fi
}%
\providecommand \@ifx [1]{%
 \ifx #1\expandafter \@firstoftwo
 \else \expandafter \@secondoftwo
 \fi
}%
\providecommand \natexlab [1]{#1}%
\providecommand \enquote  [1]{``#1''}%
\providecommand \bibnamefont  [1]{#1}%
\providecommand \bibfnamefont [1]{#1}%
\providecommand \citenamefont [1]{#1}%
\providecommand \href@noop [0]{\@secondoftwo}%
\providecommand \href [0]{\begingroup \@sanitize@url \@href}%
\providecommand \@href[1]{\@@startlink{#1}\@@href}%
\providecommand \@@href[1]{\endgroup#1\@@endlink}%
\providecommand \@sanitize@url [0]{\catcode `\\12\catcode `\$12\catcode
  `\&12\catcode `\#12\catcode `\^12\catcode `\_12\catcode `\%12\relax}%
\providecommand \@@startlink[1]{}%
\providecommand \@@endlink[0]{}%
\providecommand \url  [0]{\begingroup\@sanitize@url \@url }%
\providecommand \@url [1]{\endgroup\@href {#1}{\urlprefix }}%
\providecommand \urlprefix  [0]{URL }%
\providecommand \Eprint [0]{\href }%
\providecommand \doibase [0]{http://dx.doi.org/}%
\providecommand \selectlanguage [0]{\@gobble}%
\providecommand \bibinfo  [0]{\@secondoftwo}%
\providecommand \bibfield  [0]{\@secondoftwo}%
\providecommand \translation [1]{[#1]}%
\providecommand \BibitemOpen [0]{}%
\providecommand \bibitemStop [0]{}%
\providecommand \bibitemNoStop [0]{.\EOS\space}%
\providecommand \EOS [0]{\spacefactor3000\relax}%
\providecommand \BibitemShut  [1]{\csname bibitem#1\endcsname}%
\let\auto@bib@innerbib\@empty
\bibitem [{\citenamefont {Buzsaki}(2006)}]{buzsaki2006rhythms}%
  \BibitemOpen
  \bibfield  {author} {\bibinfo {author} {\bibfnamefont {G.}~\bibnamefont
  {Buzsaki}},\ }\href
  {https://urldefense.com/v3/__https://books.google.com/books?id=ldz58irprjYC__;!!Mih3wA!VDUnBPduahevOhOiOqYHTZZHh4H7dA_b9_L4xp0hCJOVds515-icRlSbEfQA$}
  {\emph {\bibinfo {title} {Rhythms of the Brain}}}\ (\bibinfo  {publisher}
  {Oxford University Press},\ \bibinfo {year} {2006})\BibitemShut {NoStop}%
\bibitem [{\citenamefont {Gerstner}\ \emph {et~al.}(2014)\citenamefont
  {Gerstner}, \citenamefont {Kistler}, \citenamefont {Naud},\ and\
  \citenamefont {Paninski}}]{Gerstner:2014:NDS:2635959}%
  \BibitemOpen
  \bibfield  {author} {\bibinfo {author} {\bibfnamefont {W.}~\bibnamefont
  {Gerstner}}, \bibinfo {author} {\bibfnamefont {W.~M.}\ \bibnamefont
  {Kistler}}, \bibinfo {author} {\bibfnamefont {R.}~\bibnamefont {Naud}}, \
  and\ \bibinfo {author} {\bibfnamefont {L.}~\bibnamefont {Paninski}},\
  }\href@noop {} {\emph {\bibinfo {title} {Neuronal Dynamics: From Single
  Neurons to Networks and Models of Cognition}}}\ (\bibinfo  {publisher}
  {Cambridge University Press},\ \bibinfo {address} {New York, NY, USA},\
  \bibinfo {year} {2014})\BibitemShut {NoStop}%
\bibitem [{\citenamefont {{Kuramoto}}(1975)}]{1975LNP....39..420K}%
  \BibitemOpen
  \bibfield  {author} {\bibinfo {author} {\bibfnamefont {Y.}~\bibnamefont
  {{Kuramoto}}},\ }in\ \href {\doibase 10.1007/BFb0013365} {\emph {\bibinfo
  {booktitle} {Mathematical Problems in Theoretical Physics}}},\ \bibinfo
  {series} {Lecture Notes in Physics, Berlin Springer Verlag}, Vol.~\bibinfo
  {volume} {39},\ \bibinfo {editor} {edited by\ \bibinfo {editor}
  {\bibfnamefont {H.}~\bibnamefont {{Araki}}}}\ (\bibinfo {year} {1975})\ pp.\
  \bibinfo {pages} {420--422}\BibitemShut {NoStop}%
\bibitem [{\citenamefont {Kuramoto}\ and\ \citenamefont
  {Battogtokh}(2002)}]{Kuramoto2002}%
  \BibitemOpen
  \bibfield  {author} {\bibinfo {author} {\bibfnamefont {Y.}~\bibnamefont
  {Kuramoto}}\ and\ \bibinfo {author} {\bibfnamefont {D.}~\bibnamefont
  {Battogtokh}},\ }\href@noop {} {\bibfield  {journal} {\bibinfo  {journal}
  {Nonlinear Phenom. Complex Syst.}\ }\textbf {\bibinfo {volume} {5}},\
  \bibinfo {pages} {380} (\bibinfo {year} {2002})}\BibitemShut {NoStop}%
\bibitem [{\citenamefont {Kuramoto}(2002)}]{Kuramoto2003}%
  \BibitemOpen
  \bibfield  {author} {\bibinfo {author} {\bibfnamefont {Y.}~\bibnamefont
  {Kuramoto}},\ }in\ \href@noop {} {\emph {\bibinfo {booktitle} {{N}onlinear
  {D}ynamics and {C}haos: {W}here do we go from here?}}},\ \bibinfo {editor}
  {edited by\ \bibinfo {editor} {\bibfnamefont {J.}~\bibnamefont {Hogan}},
  \bibinfo {editor} {\bibfnamefont {A.}~\bibnamefont {Krauskopf}}, \bibinfo
  {editor} {\bibfnamefont {M.}~\bibnamefont {di~Bernado}}, \bibinfo {editor}
  {\bibfnamefont {R.}~\bibnamefont {Wilson}}, \bibinfo {editor} {\bibfnamefont
  {H.}~\bibnamefont {Osinga}}, \bibinfo {editor} {\bibfnamefont
  {M.}~\bibnamefont {Homer}}, \ and\ \bibinfo {editor} {\bibfnamefont
  {A.}~\bibnamefont {Champneys}}}\ (\bibinfo  {publisher} {CRC Press},\
  \bibinfo {year} {2002})\ pp.\ \bibinfo {pages} {209--227}\BibitemShut
  {NoStop}%
\bibitem [{\citenamefont {{Abrams}}\ and\ \citenamefont
  {{Strogatz}}(2004)}]{2004PhRvL..93q4102A}%
  \BibitemOpen
  \bibfield  {author} {\bibinfo {author} {\bibfnamefont {D.~M.}\ \bibnamefont
  {{Abrams}}}\ and\ \bibinfo {author} {\bibfnamefont {S.~H.}\ \bibnamefont
  {{Strogatz}}},\ }\href {\doibase 10.1103/PhysRevLett.93.174102} {\bibfield
  {journal} {\bibinfo  {journal} {Physical Review Letters}\ }\textbf {\bibinfo
  {volume} {93}},\ \bibinfo {eid} {174102} (\bibinfo {year} {2004})},\ \Eprint
  {http://arxiv.org/abs/nlin/0407045} {nlin/0407045} \BibitemShut {NoStop}%
\bibitem [{\citenamefont {{Acebr{\'o}n}}\ \emph {et~al.}(2005)\citenamefont
  {{Acebr{\'o}n}}, \citenamefont {{Bonilla}}, \citenamefont {{P{\'e}rez
  Vicente}}, \citenamefont {{Ritort}},\ and\ \citenamefont
  {{Spigler}}}]{2005RvMP...77..137A}%
  \BibitemOpen
  \bibfield  {author} {\bibinfo {author} {\bibfnamefont {J.~A.}\ \bibnamefont
  {{Acebr{\'o}n}}}, \bibinfo {author} {\bibfnamefont {L.~L.}\ \bibnamefont
  {{Bonilla}}}, \bibinfo {author} {\bibfnamefont {C.~J.}\ \bibnamefont
  {{P{\'e}rez Vicente}}}, \bibinfo {author} {\bibfnamefont {F.}~\bibnamefont
  {{Ritort}}}, \ and\ \bibinfo {author} {\bibfnamefont {R.}~\bibnamefont
  {{Spigler}}},\ }\href {\doibase 10.1103/RevModPhys.77.137} {\bibfield
  {journal} {\bibinfo  {journal} {Reviews of Modern Physics}\ }\textbf
  {\bibinfo {volume} {77}},\ \bibinfo {pages} {137} (\bibinfo {year}
  {2005})}\BibitemShut {NoStop}%
\bibitem [{\citenamefont {Escaff}\ and\ \citenamefont
  {Delpiano}(2020)}]{pmid32872818}%
  \BibitemOpen
  \bibfield  {author} {\bibinfo {author} {\bibfnamefont {D.}~\bibnamefont
  {Escaff}}\ and\ \bibinfo {author} {\bibfnamefont {R.}~\bibnamefont
  {Delpiano}},\ }\href@noop {} {\bibfield  {journal} {\bibinfo  {journal}
  {Chaos}\ }\textbf {\bibinfo {volume} {30}},\ \bibinfo {pages} {083137}
  (\bibinfo {year} {2020})}\BibitemShut {NoStop}%
\bibitem [{\citenamefont {Wu}\ \emph {et~al.}(2018)\citenamefont {Wu},
  \citenamefont {Kang}, \citenamefont {Liu},\ and\ \citenamefont
  {Dhamala}}]{pmid30341395}%
  \BibitemOpen
  \bibfield  {author} {\bibinfo {author} {\bibfnamefont {H.}~\bibnamefont
  {Wu}}, \bibinfo {author} {\bibfnamefont {L.}~\bibnamefont {Kang}}, \bibinfo
  {author} {\bibfnamefont {Z.}~\bibnamefont {Liu}}, \ and\ \bibinfo {author}
  {\bibfnamefont {M.}~\bibnamefont {Dhamala}},\ }\href@noop {} {\bibfield
  {journal} {\bibinfo  {journal} {Sci Rep}\ }\textbf {\bibinfo {volume} {8}},\
  \bibinfo {pages} {15521} (\bibinfo {year} {2018})}\BibitemShut {NoStop}%
\bibitem [{\citenamefont {Hodgkin}\ and\ \citenamefont
  {Huxley}(1952)}]{pmid12991237}%
  \BibitemOpen
  \bibfield  {author} {\bibinfo {author} {\bibfnamefont {A.~L.}\ \bibnamefont
  {Hodgkin}}\ and\ \bibinfo {author} {\bibfnamefont {A.~F.}\ \bibnamefont
  {Huxley}},\ }\href@noop {} {\bibfield  {journal} {\bibinfo  {journal} {J.
  Physiol. (Lond.)}\ }\textbf {\bibinfo {volume} {117}},\ \bibinfo {pages}
  {500} (\bibinfo {year} {1952})}\BibitemShut {NoStop}%
\bibitem [{\citenamefont {Fitzhugh}(1961)}]{pmid19431309}%
  \BibitemOpen
  \bibfield  {author} {\bibinfo {author} {\bibfnamefont {R.}~\bibnamefont
  {Fitzhugh}},\ }\href@noop {} {\bibfield  {journal} {\bibinfo  {journal}
  {Biophys. J.}\ }\textbf {\bibinfo {volume} {1}},\ \bibinfo {pages} {445}
  (\bibinfo {year} {1961})}\BibitemShut {NoStop}%
\bibitem [{\citenamefont {Nagumo}\ \emph {et~al.}(1962)\citenamefont {Nagumo},
  \citenamefont {Arimoto},\ and\ \citenamefont {Yoshizawa}}]{Nagumo1962}%
  \BibitemOpen
  \bibfield  {author} {\bibinfo {author} {\bibfnamefont {J.}~\bibnamefont
  {Nagumo}}, \bibinfo {author} {\bibfnamefont {S.}~\bibnamefont {Arimoto}}, \
  and\ \bibinfo {author} {\bibfnamefont {S.}~\bibnamefont {Yoshizawa}},\ }\href
  {\doibase 10.1109/jrproc.1962.288235} {\bibfield  {journal} {\bibinfo
  {journal} {Proceedings of the {IRE}}\ }\textbf {\bibinfo {volume} {50}},\
  \bibinfo {pages} {2061} (\bibinfo {year} {1962})}\BibitemShut {NoStop}%
\bibitem [{\citenamefont {Morris}\ and\ \citenamefont
  {Lecar}(1981)}]{pmid7260316}%
  \BibitemOpen
  \bibfield  {author} {\bibinfo {author} {\bibfnamefont {C.}~\bibnamefont
  {Morris}}\ and\ \bibinfo {author} {\bibfnamefont {H.}~\bibnamefont {Lecar}},\
  }\href@noop {} {\bibfield  {journal} {\bibinfo  {journal} {Biophys. J.}\
  }\textbf {\bibinfo {volume} {35}},\ \bibinfo {pages} {193} (\bibinfo {year}
  {1981})}\BibitemShut {NoStop}%
\bibitem [{\citenamefont {Izhikevich}(2003)}]{pmid18244602}%
  \BibitemOpen
  \bibfield  {author} {\bibinfo {author} {\bibfnamefont {E.~M.}\ \bibnamefont
  {Izhikevich}},\ }\href@noop {} {\bibfield  {journal} {\bibinfo  {journal}
  {IEEE Trans Neural Netw}\ }\textbf {\bibinfo {volume} {14}},\ \bibinfo
  {pages} {1569} (\bibinfo {year} {2003})}\BibitemShut {NoStop}%
\bibitem [{\citenamefont {Kulkarni}\ \emph {et~al.}(2020)\citenamefont
  {Kulkarni}, \citenamefont {Ranft},\ and\ \citenamefont
  {Hakim}}]{pmid33192427}%
  \BibitemOpen
  \bibfield  {author} {\bibinfo {author} {\bibfnamefont {A.}~\bibnamefont
  {Kulkarni}}, \bibinfo {author} {\bibfnamefont {J.}~\bibnamefont {Ranft}}, \
  and\ \bibinfo {author} {\bibfnamefont {V.}~\bibnamefont {Hakim}},\
  }\href@noop {} {\bibfield  {journal} {\bibinfo  {journal} {Front Comput
  Neurosci}\ }\textbf {\bibinfo {volume} {14}},\ \bibinfo {pages} {569644}
  (\bibinfo {year} {2020})}\BibitemShut {NoStop}%
\bibitem [{\citenamefont {Kim}\ and\ \citenamefont
  {Sejnowski}(2021)}]{pmid33288909}%
  \BibitemOpen
  \bibfield  {author} {\bibinfo {author} {\bibfnamefont {R.}~\bibnamefont
  {Kim}}\ and\ \bibinfo {author} {\bibfnamefont {T.~J.}\ \bibnamefont
  {Sejnowski}},\ }\href@noop {} {\bibfield  {journal} {\bibinfo  {journal} {Nat
  Neurosci}\ }\textbf {\bibinfo {volume} {24}},\ \bibinfo {pages} {129}
  (\bibinfo {year} {2021})}\BibitemShut {NoStop}%
\bibitem [{\citenamefont {Arnulfo}\ \emph {et~al.}(2020)\citenamefont
  {Arnulfo}, \citenamefont {Wang}, \citenamefont {Myrov}, \citenamefont
  {Toselli}, \citenamefont {Hirvonen}, \citenamefont {Fato}, \citenamefont
  {Nobili}, \citenamefont {Cardinale}, \citenamefont {Rubino}, \citenamefont
  {Zhigalov}, \citenamefont {Palva},\ and\ \citenamefont
  {Palva}}]{pmid33097714}%
  \BibitemOpen
  \bibfield  {author} {\bibinfo {author} {\bibfnamefont {G.}~\bibnamefont
  {Arnulfo}}, \bibinfo {author} {\bibfnamefont {S.~H.}\ \bibnamefont {Wang}},
  \bibinfo {author} {\bibfnamefont {V.}~\bibnamefont {Myrov}}, \bibinfo
  {author} {\bibfnamefont {B.}~\bibnamefont {Toselli}}, \bibinfo {author}
  {\bibfnamefont {J.}~\bibnamefont {Hirvonen}}, \bibinfo {author}
  {\bibfnamefont {M.~M.}\ \bibnamefont {Fato}}, \bibinfo {author}
  {\bibfnamefont {L.}~\bibnamefont {Nobili}}, \bibinfo {author} {\bibfnamefont
  {F.}~\bibnamefont {Cardinale}}, \bibinfo {author} {\bibfnamefont
  {A.}~\bibnamefont {Rubino}}, \bibinfo {author} {\bibfnamefont
  {A.}~\bibnamefont {Zhigalov}}, \bibinfo {author} {\bibfnamefont
  {S.}~\bibnamefont {Palva}}, \ and\ \bibinfo {author} {\bibfnamefont {J.~M.}\
  \bibnamefont {Palva}},\ }\href@noop {} {\bibfield  {journal} {\bibinfo
  {journal} {Nat Commun}\ }\textbf {\bibinfo {volume} {11}},\ \bibinfo {pages}
  {5363} (\bibinfo {year} {2020})}\BibitemShut {NoStop}%
\bibitem [{\citenamefont {Gold}\ \emph {et~al.}(2009)\citenamefont {Gold},
  \citenamefont {Girardin}, \citenamefont {Martin},\ and\ \citenamefont
  {Koch}}]{pmid19793873}%
  \BibitemOpen
  \bibfield  {author} {\bibinfo {author} {\bibfnamefont {C.}~\bibnamefont
  {Gold}}, \bibinfo {author} {\bibfnamefont {C.~C.}\ \bibnamefont {Girardin}},
  \bibinfo {author} {\bibfnamefont {K.~A.}\ \bibnamefont {Martin}}, \ and\
  \bibinfo {author} {\bibfnamefont {C.}~\bibnamefont {Koch}},\ }\href@noop {}
  {\bibfield  {journal} {\bibinfo  {journal} {J Neurophysiol}\ }\textbf
  {\bibinfo {volume} {102}},\ \bibinfo {pages} {3340} (\bibinfo {year}
  {2009})}\BibitemShut {NoStop}%
\bibitem [{\citenamefont {Galinsky}\ and\ \citenamefont
  {Frank}(2020{\natexlab{a}})}]{Galinsky:2020a}%
  \BibitemOpen
  \bibfield  {author} {\bibinfo {author} {\bibfnamefont {V.~L.}\ \bibnamefont
  {Galinsky}}\ and\ \bibinfo {author} {\bibfnamefont {L.~R.}\ \bibnamefont
  {Frank}},\ }\href@noop {} {\bibfield  {journal} {\bibinfo  {journal}
  {Physical Review Research}\ }\textbf {\bibinfo {volume} {2}},\ \bibinfo
  {pages} {023061} (\bibinfo {year} {2020}{\natexlab{a}})}\BibitemShut
  {NoStop}%
\bibitem [{\citenamefont {Galinsky}\ and\ \citenamefont
  {Frank}(2020{\natexlab{b}})}]{Galinsky:2020b}%
  \BibitemOpen
  \bibfield  {author} {\bibinfo {author} {\bibfnamefont {V.~L.}\ \bibnamefont
  {Galinsky}}\ and\ \bibinfo {author} {\bibfnamefont {L.~R.}\ \bibnamefont
  {Frank}},\ }\href@noop {} {\bibfield  {journal} {\bibinfo  {journal} {J. of
  Cognitive Neurosci}\ }\textbf {\bibinfo {volume} {32}},\ \bibinfo {pages}
  {2178} (\bibinfo {year} {2020}{\natexlab{b}})}\BibitemShut {NoStop}%
\bibitem [{\citenamefont {Zakharov}\ \emph {et~al.}(1992)\citenamefont
  {Zakharov}, \citenamefont {L'vov},\ and\ \citenamefont
  {Falkovich}}]{book:971420}%
  \BibitemOpen
  \bibfield  {author} {\bibinfo {author} {\bibfnamefont {V.~E.}\ \bibnamefont
  {Zakharov}}, \bibinfo {author} {\bibfnamefont {V.~S.}\ \bibnamefont {L'vov}},
  \ and\ \bibinfo {author} {\bibfnamefont {G.}~\bibnamefont {Falkovich}},\
  }\href@noop {} {\emph {\bibinfo {title} {Kolmogorov Spectra of Turbulence I:
  Wave Turbulence}}},\ \bibinfo {edition} {1st}\ ed.,\ Springer Series in
  Nonlinear Dynamics\ (\bibinfo  {publisher} {Springer-Verlag Berlin
  Heidelberg},\ \bibinfo {year} {1992})\BibitemShut {NoStop}%
\bibitem [{\citenamefont {Nazarenko}(2011)}]{book:787941}%
  \BibitemOpen
  \bibfield  {author} {\bibinfo {author} {\bibfnamefont {S.}~\bibnamefont
  {Nazarenko}},\ }\href@noop {} {\emph {\bibinfo {title} {Wave Turbulence}}},\
  \bibinfo {edition} {1st}\ ed.,\ Lecture Notes in Physics 825\ (\bibinfo
  {publisher} {Springer-Verlag Berlin Heidelberg},\ \bibinfo {year}
  {2011})\BibitemShut {NoStop}%
\bibitem [{Syn()}]{Sync-Supp}%
  \BibitemOpen
  \href@noop {} {}\bibinfo {note} {See Supplemental Material 
  for Wolfram Mathematica notebook with parameters and
  examples of numerical integration of systems Eqs.~\cref{eq:A0,eq:B0,eq:AM0,eq:BM0}
  used for the plots of \cref{fig:spiking,fig:f2}.}\BibitemShut {Stop}%
\end{thebibliography}
%

\end{document}